\documentclass{rspublic}
\usepackage{psfig}

\def\apj{{\it Astrophys. J.}}
\def\nat{{\it Nature}}
\def\mnras{{\it Mon. Not. R. Astr. Soc.}}
\def\aap{{\it Astr. \& Astrophys.}}
\def\apjl{{\it Astrophys. J. Lett.}}
\def\pasj{{\it Publ. Astr. Soc. Japan}}
\def\gsim{\mathrel{\hbox{\rlap{\lower.55ex \hbox {$\sim$}}\kern-.0em
\raise.4ex \hbox{$>$}}}}

\def\etal{{\it et al.}}
\def\swift{{\it Swift}}

\begin{document}

\title[GRBs at high redshift]
{Observations of GRBs at High Redshift}

\author[N.~R. Tanvir, P. Jakobsson]{Nial R. Tanvir$^1$, P\'all Jakobsson$^2$}

\affiliation{$^1$Department of Physics and Astronomy, 
University of Leicester, Leicester LE1 7RH, UK;
$^2$Centre for Astrophysics Research,
University of Hertfordshire, Hatfield, Herts, AL10 9AB, UK}

\label{firstpage}

\maketitle

\begin{abstract}{Gamma-ray bursts; Host galaxies; Cosmic reionization}
The extreme luminosity of gamma-ray bursts and their
afterglows means they are detectable, in principle, to 
very high redshifts.  Although the redshift distribution
of GRBs is difficult to determine, due to incompleteness
of present samples, we argue that for {\it Swift}-detected bursts
the median redshift is between 2.5 and 3, with a few
percent likely at $z>6$.
Thus, GRBs are potentially powerful probes of the
era of reionization, and the sources responsible for it.
Moreover, it seems likely that they can provide 
constraints on the star formation history of the universe, and
may also help in the determination of the cosmological parameters.
\end{abstract}

\section{Introduction}

Long-duration gamma-ray bursts (GRBs) are becoming powerful probes of
the distant universe.  The overriding asset of GRBs is extreme
luminosity over a very broad wavelength range, which in principle
allows them to be seen and studied to very high redshift, $z\sim20$
(Lamb \& Reichart 2000).  They can be detected in
gamma-rays even in the presence of high column densities of
intervening material, removing one potential source of incompleteness
bias.  Furthermore, having stellar progenitors, they can
be detected irrespective of the luminosity of the star-forming
host itself.  The
drawback of GRBs is, of course, their rarity, and even in the {\it
Swift} era it continues to be a time-consuming business building up
statistically useful samples.

The highest redshift found to-date is $z=6.30$
for GRB 050904 (Haislip \etal\ 2006; Kawai \etal\ 2006).
This compares well with the most distant galaxy with a
confirmed spectroscopic redshift of $z=6.95$
(Iye \etal\ 2006).
The number density of bright galaxies (and quasars) decreases
rapidly at early times, as expected in hierarchical growth
of structure, making them increasingly 
rare and hard to detect at $z>6$
(cf. Reed \etal\ 2003),
and hence the searches for even higher redshift GRBs all
the more important.

In this review we first consider the crucial question of
the redshift distribution of GRBs, and what we can say about the numbers 
likely to be found at very high redshifts.
This includes consideration of the importance of ``dark bursts'':
those with very faint or undetected optical afterglows.

We then outline some of the scientific programmes proposed or
under way to use GRBs to illuminate a number of cosmological
questions, particularly considering the role of GRBs as 
star formation indicators, and as a means of studying
high-redshift galaxies and their environments.

\section{The Redshift Distribution of GRBs}

Various authors have attempted to predict the redshift
distribution of GRBs.  The ingredients of such models are
basically (a) a parametrization or prediction
of the star-formation history
of the universe; combined with (b) some mapping from star-formation
rate to GRB rate and luminosity; and finally (c) convolving
with a selection function dependent on the instrument
used for initial detection.  The star-formation history
of the universe is already uncertain,
particularly beyond $z=5$, which of course is one of
the motivations for pursuing GRB studies, as discussed below.
Mapping star formation rate onto GRB rate, 
essentially an attempt to guess the sensitivity
of GRB rate and/or luminosity on factors such as metallicity,
and indeed the metallicity distribution amongst the stellar
populations present at a given redshift, is currently
a matter of educated guess-work.
Even the selection function of {\it Swift}/BAT is non-trivial
since the detectability of a given GRB depends on its spectrum
and photon time history, along with instrumental factors such as position
in field of view and the state of the (evolving) detection
algorithm.

Despite all these provisos, it is important to have some 
predictions for plausible rates.  Notable recent attempts 
include:
Guetta, Piran \& Waxman (2005) explore a variety
of different star-formation rate histories and
GRB luminosity functions;
Natarajan \etal\ (2005) additionally incorporate a simple
prescription for a low-metallicity preference
for GRBs; Yoon \& Langer (2005) take a more sophisticated
approach to the metallicity question by explicitly
identifying stellar evolution models which naturally
lead to collapsar progenitors;
Bromm \& Loeb (2006) further consider the
possibility of a population III contribution to the high-$z$
GRB rate. 

The observed redshift distribution should also be treated with caution
because it is susceptible to further
important selection effects.  For example,
those GRBs with bright optical afterglows are much more likely
to have a redshift measured than those for which the optical
afterglow is faint
(cosmic time-dilation does help here to some extent, since
at fixed observer-time we see an earlier, and usually
intrinsically brighter, phase of the afterglow).

In fact, of all the GRBs which have been reasonably well-localised,
less than 40\% have had direct redshift measurements, making
them a highly incomplete sample.
A few redshifts have been measured for
``dark'' bursts, but only when the GRB is pinned down well enough by
its X-ray or radio position to identify a likely host, and the host  
itself is a sufficiently bright to obtain a spectroscopic redshift.
However, it is important to remember that
in many cases
the lack of an optical afterglow may be blamed on poor
positioning of the burst, for example at low Galactic-latitude,
or being too close to the Sun or bright Moon for deep followup.

In an effort to improve this situation, Jakobsson \etal\ (2006)
defined a subset of all GRBs {\em well-placed for optical observation}.
The selection criteria were that the burst should have an X-ray
position made public within 12 hours, the Galactic foreground
be low $A_V<0.5$, the burst be $>55^{\circ}$ from the Sun, and
not at a polar declination, $|{\rm dec}|<70^{\circ}$.
Imposing these restrictions reduces the GRB sample size, but
greatly increases the redshift completeness of those samples.
In fact, Jakobsson \etal\ (2006) were able to show that
the median redshift of \swift\ discovered GRBs is considerably 
higher than that of pre-\swift\ GRBs.
An updated illustration of this difference is shown by the red
and blue lines in figure~1.

\begin{figure}
 \psfig{figure=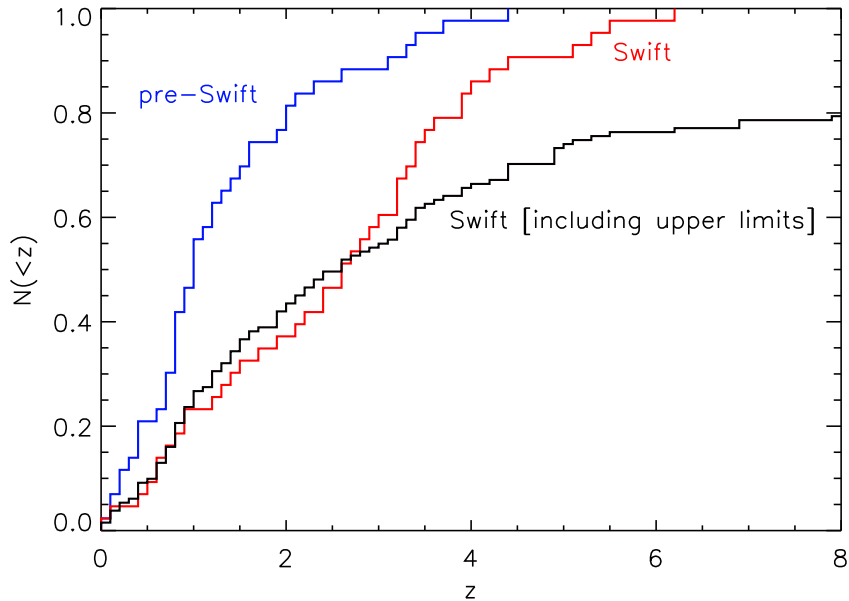,width=11.95cm}
\caption{Cumulative histograms of redshifts obtained for
uniformly-selected samples of long-duration bursts as
of end November 2006.  The pre-{\it
Swift} sample (blue) had a relatively low median redshift around
$z=1$.  For the {\it Swift} GRBs (selected following the criteria of
Jakobsson \etal\ 2006) the two lines are (a) an incomplete sub-sample
with only those 43 bursts with well-determined redshifts (red); and, (b) the
full sample of 83, including 40 bursts without redshifts (ie. generally
optically fainter) placed in the distribution according to their 
upper redshift limits based on photometry
(black).  The true, optically-unbiassed {\it Swift} redshift
distribution is likely, therefore, to be between these two lines.}
\end{figure}

Unfortunately, even with the above restrictions, only 50--60\% 
of the  \swift\ GRB sample
have good redshifts (spectroscopic or photometric with many bands).
However, many of the remaining bursts did at least have detections
in UV, optical or nIR bands which can be used to place an {\em upper
limit} to their redshift. In particular, when detected in the UV
by UVOT, the upper limits can be relatively low.
Indeed less than 20\% of the sample have no 
constraint on their redshift.

To visualise the maximum possible effect of redshift incompleteness we
also plot in figure~1 a black  line which includes now {\em all}
the bursts satisfying the selection criteria, but placing those
with no firm redshift at the {\em maximum} redshift they can have
given their bluest photometric detection.
Reality is likely to lie somewhere between the red and black curves,
which is consistent with the Bromm \& Loeb (2006) prediction that
$\sim$10\% of {\it Swift} GRBs should lie beyond $z=5$.

\section{GRBs as a Means of Studying High-$z$ Galaxies}

Most of the methods used to find and study high-redshift galaxies
rely on detecting the galaxy in some waveband.
In particular, in recent years Lyman-break galaxies (LBGs), 
identified via their optical (rest-frame UV) colours (Steidel \etal\ 2003);
submm galaxies (Ivison \etal\ 2005); 
and Lyman-$\alpha$ emission line galaxies
(Malhotra \& Rhoads 2006; Iye \etal\ 2006), 
have been central to our understanding of the high-$z$
galaxy population.
However, since these techniques all rely on the galaxy being
luminous enough to make it into the samples, they are biased towards
the bright end of the galaxy LF in whichever waveband the search
is performed.

Quasar absorption lines can be used to locate galaxies based on
absorbing column rather than luminosity, but it can be
problematic to identify the counterpart in emission against
the bright quasar point source.

GRBs on the other hand, have stellar progenitors, and therefore
select galaxies independently of their luminosities.  Their
afterglow absorption line spectra give redshifts, and indeed
chemical and dynamical information about the host's 
interstellar medium, even
for extremely faint galaxies.
Furthermore, once the afterglow has faded, the host can
be studied directly.
A good example of the power of GRBs as a means of selecting
high-redshift galaxies is that of GRB 020124, which was
undetected in a deep {\it HST} pointing to $R\approx29.5$
(Berger \etal\ 2002), and
yet was found to be a damped Lyman-$\alpha$ system at 
$z=3.20$ through followup of its afterglow (Hjorth \etal\ 2003a).
Vreeswijk \etal\ (2004) made a pioneering high-S/N observation
of another notable burst with a faint host galaxy, that of GRB 030323
with $R_{AB}\approx28$, 
showing it to have a metallicity only a few percent of solar.

Studies of host samples continue to be troubled by the
incompleteness of spectroscopy.  However, despite that,
interesting comparisons have been made between GRB hosts
and other populations.
In particular, Fruchter \etal\ (2006) recently
compared GRB hosts to the hosts of core-collapse supernovae
in the same redshift range, and showed them to have
quite different morphologies and luminosities on the
average.  Since both are thought to have massive star
progenitors, this is surprising, although it confirms
the long-noted fact that many GRB hosts are rather small,
irregular galaxies, and could be a consequence of
a metallicity dependence of GRB properties.


\section{GRBs as Tracers of Star Formation}

Since GRBs are produced by massive stars (eg. Hjorth \etal\ 2003b), 
which have
short life-times, one would expect
the rate of GRBs is proportional to the
massive star formation rate at any given epoch
(eg. Wijers \etal\ 1998).
Assuming a universal IMF (in common with most other
SFR estimation techniques) allows us to infer a
total star formation rate history.

Advantages of GRBs as star formation indicators are
that they are very bright and can be seen through high columns
of gas and dust.  Furthermore, as discussed above, we can
count GRBs as a function of redshift even when their hosts
are too faint to have appeared in any photometric census
of star formation.

Our hope then is that if GRBs can be shown to be an unbiassed
tracer of star formation, then the redshift distribution of GRBs
can in principle be inverted to give the global star formation rate 
history.  The proportion of star formation occuring in different
populations of galaxies, should be reflected in the proportions
of such galaxies amongst the GRB hosts.

This hypothesis can be tested by looking at the distribution
of star-forming properties of GRB hosts.
In the optical/UV, Jakobsson \etal\ (2005) found that a small
sample of GRB hosts with $z\gsim2$ 
had a luminosity function consistent
with being a star-formation-weighted Lyman-break galaxy
luminosity function.
Since a large proportion of high redshift star formation
is thought to be dust-obscured, it is of particular
interest to investigate the fIR/submm/radio properties of GRB
hosts.  Early results showed that some hosts were indeed
detectable in submm or radio 
(Frail \etal\ 2002; Barnard \etal\ 2003; Berger \etal\ 2003b).
Included amongst the target sample were a number of hosts
of ``dark'' bursts (identified via accurate X-ray locations),
helping mitigate against an optical bias.

However,  subsequently it has become clear that the
numbers are significantly below predictions based on the 
expected high proportion of global star formation taking
place in ultra-luminous dusty galaxies 
(eg. Tanvir \etal\ 2004; Priddey \etal\ 2006;
Le Floc'h \etal\ 2006).
The conclusion of these studies is that whilst GRB hosts
are in general actively star-forming, very few are intensively
star-forming ULIRG-like galaxies.  
Once again, a plausible explanation is that GRBs are 
selecting smaller, lower-metallicity galaxies
(cf. Fynbo \etal\ 2003), which would
make them less useful as a probe of all star formation,
but possibly increasingly useful as a tracer of higher
redshift star formation.

\section{GRBs as a means of Studying the Era of Reionization}

In conventional cosmology, 
after recombination the universe remained neutral until the first
collapsed sources began to emit UV radiation.  This radiation 
ionized a region around each source, and these regions eventually
grew and merged to form a nearly fully ionized intergalactic medium
by a redshift of about $z=6$.  This is known from spectroscopy of
high-redshift quasars which show that the neutral fraction was
low (and dropping) at these redshifts.

Measurements of the electron-scattering optical depth by
the Microwave Anisotropy Probe
(WMAP) observations of the microwave background, 
however, indicate that reionization 
was substantially under way at earlier times, around $z=11$
(Page \etal\ 2006; Spergel \etal\ 2006).
The epoch of the very earliest collapsed sources,
and the detailed time history of reionization,
which may have proceeded in a slow continuous way or
in separate pop III and pop II phases, remain
open questions (eg. Furlanetto \& Loeb 2005).

There is considerable interest in the nature of the
first objects, and the phase change that they brought about.  
However, probing
further with QSOs becomes difficult because of the rapidly
diminishing number density of bright QSOs beyond $z=6$.
QSO spectra are also difficult to analyse due to the bright
emission lines
and the substantial ``proximity'' 
effect that a bright QSO has on the surrounding intergalactic
space.
GRBs, on the other hand, have stellar progenitors, and so may
well be frequent in the high-$z$ universe.  They tend to reside in
small galaxies, with little proximity effect, and they have power-law
continua against which it is relatively easy, in principle, to measure
absorption features
(eg. Mesinger, Haiman \& Cen 2004).

The difficulty with GRBs is, of course, their transience and 
rarity.  It is therefore essential that any high-$z$ bursts that
are discovered are identified as such as soon as possible, and
followed up with nIR spectroscopy while the afterglow is still
bright.  In the case of GRB 050904 the afterglow was already
very faint when the Subaru spectrum was obtained, but
an estimate of the neutral fraction was still possible
(Totani \etal\ 2006; and Kawai \etal\ 2006).

\section{GRBs to Measure Cosmological Parameters}

Given the diverse behaviour of GRBs, particularly of their
prompt emission, 
the prospects for their use as distance
indicators would not seem promissing at first sight.
Nonetheless
Berger, Kulkarni \& Frail (2003)
showed that after correcting for beaming
the majority of bursts seemed to have a ``standard reservoir''
of energy that they released. 
Several other authors have also explored the correlation
between GRB luminosity (or energy) and other
parameters that are independent (or partially
independent) of distance.
Notably Ghirlanda, Ghisellini \& Lazzati (2004)
found a remarkably tight relation 
collimation-corrected  energy and the peak energy of
the $\nu$F$_{\nu}$ prompt spectrum.

The advantage of GRBs in this area is that they are
found to considerably higher redshifts than SNeIa,
and so provide a complementary constraint on 
cosmological world-models.  The main drawback 
at the present is that there is only a small
and rather inhomogeneous sample of bursts available
which must be used to both calibrate the relation
and provide cosmological constraints.
For further discussion
see Ghirlanda in this volume.

\section{Conclusions}

We have seen that long-duration GRBs hold considerable
promise as probes of the high-redshift universe.
As a final illustration, in figure~2 we show the 
history of the most distant known quasar, galaxy
and GRB over the past $\sim$50 years.
Although only a relative new-comer to this game, GRBs
have rapidly become competetive, and there are
reasons to hope that with {\it Swift, GLAST} and other
satellites providing hundreds of localisations over
the next few years, that they may become the method
of choice for studies of the earliest era of structure
formation.

\begin{figure}
 \psfig{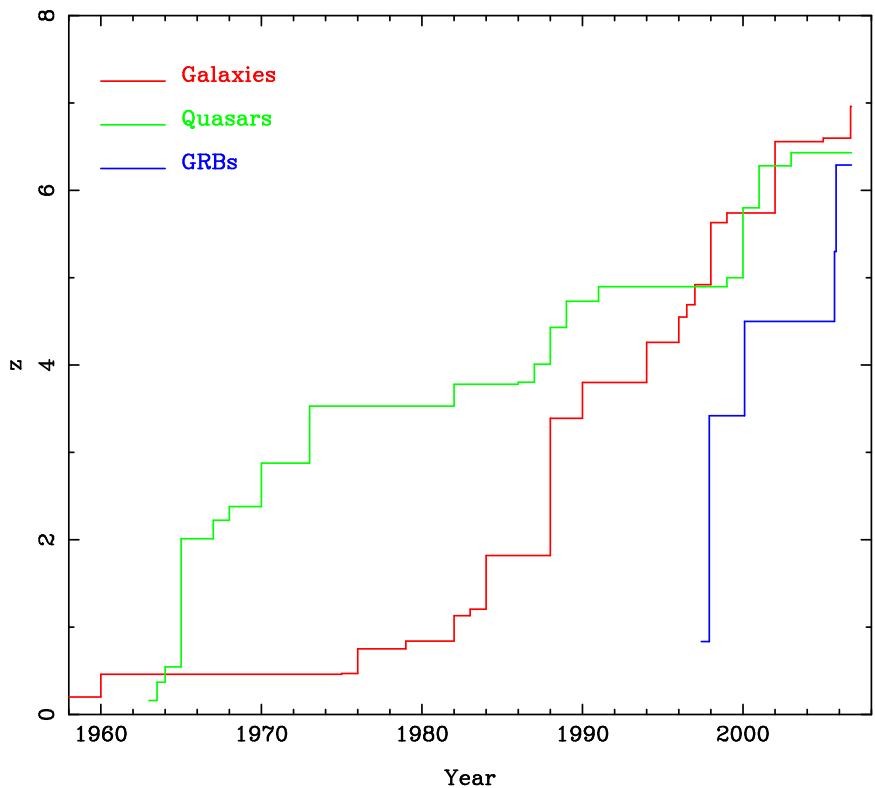}
\caption{
A representation of the historical increase in 
the highest known redshift for quasars and galaxies
as compared to GRBs.
}
\end{figure}

\begin{acknowledgements}
 We thank R. McMahon for providing many of the references
 which went into making figure~2, and 
 A. Levan and R. Priddey for useful discussions.
\end{acknowledgements}


\label{lastpage}
\end{document}